\documentclass[floats,prd,aps,showpacs,amssymb,nofootinbib,onecolumn]{revtex4}
\usepackage{amssymb}
\usepackage{graphics}
\usepackage{amsmath}
\usepackage{amsfonts}
\usepackage{bm}
\usepackage[]{latexsym}

\newcommand{\be}{\begin{equation}}\newcommand{\ee}{\end{equation}}
\newcommand{\bea}{\begin{eqnarray}}\newcommand{\eea}{\end{eqnarray}}
\newcommand{\brr}{\begin{array}}\newcommand{\err}{\end{array}}
\newcommand{\bit}{\begin{itemize}}\newcommand{\eit}{\end{itemize}}
\newcommand{\ben}{\begin{enumerate}}\newcommand{\een}{\end{enumerate}}

\newcommand{\ba}{\begin{array}}
\newcommand{\ea}{\end{array}}

\def\al{\alpha}

\def\1{{_{1}}}\def\2{{_{2}}}

\def\noHe0{:\;\!\!\;\!\!:H_e(0):\;\!\!\;\!\!:}
\def\noHm0{:\;\!\!\;\!\!:H_\mu(0):\;\!\!\;\!\!:}
\def\al{\alpha}

\def\1{{_{1}}}\def\2{{_{2}}}

\def\bogu{U_k}
\def\bogv{V_k}

\begin{document}
\title{Neutrino mixing and General Covariance in the inverse $\beta$ decay}

\author{M.Blasone\footnote{blasone@sa.infn.it}$^{\hspace{0.3mm}1,2}$, G.Lambiase\footnote{lambiase@sa.infn.it}$^{\hspace{0.3mm}1,2}$, G.G.Luciano\footnote{gluciano@sa.infn.it}$^{\hspace{0.3mm}1,2}$ 
and L.Petruzziello\footnote{lpetruzziello@na.infn.it}$^{\hspace{0.3mm}1,2}$} \affiliation
{$^1$Dipartimento di Fisica, Universit\`a di Salerno, Via Giovanni Paolo II, 132 I-84084 Fisciano (SA), Italy.\\ $^2$INFN, Sezione di Napoli, Gruppo collegato di Salerno, Italy.}

\date{\today}
\def\be{\begin{equation}}
\def\ee{\end{equation}}
\def\al{\alpha}
\def\bea{\begin{eqnarray}}
\def\eea{\end{eqnarray}}

\begin{abstract}
We review recent developments on the r\^ole of neutrino mixing in the inverse $\beta$ decay of accelerated protons. We
show that  calculations in the inertial and comoving frames agree -- thus preserving General Covariance -- only when taking neutrino asymptotic states to be flavor (rather than mass) eigenstates. Our conclusions are valid in the approximation in which Pontecorvo states are correctly representing neutrino flavor states. We speculate about the general case involving exact flavor states and finally comment on other approaches recently appeared in literature. 
\end{abstract}

\vskip -1.0 truecm 

\maketitle

\section{Introduction}
\setcounter{equation}{0}
The Unruh effect~\cite{Unruh:1976db} is one of the most important achievements of Quantum Field Theory (QFT) in curved backgrounds. Since its discovery, it was clear that a direct evidence of this phenomenon would have required a tremendous effort, due to the difficulty in detecting the Unruh temperature
\be
T_U=\frac{\hbar\,a}{2\,\pi\,c\,k_B},
\ee
which appears to be extremely small even for huge accelerations $a$. 

Even though actual experiments cannot provide a satisfactory outcome, from the theoretical point of view the Unruh effect turns out be indispensable to maintain the internal consistency of QFT. Its existence, indeed, was found to be mandatory first for clarifying the apparently controversial problem of the QED bremsstrahlung~\cite{Higuchi:1992td}, and then for 
preserving the equality of the inverse $\beta$ decay rates of accelerated proton in the inertial and comoving frames~\cite{Matsas:1999jx}, thus guaranteeing the general covariance of the underlying theoretical framework. 

An interesting development along this line was provided by the analysis of the inverse $\beta$ decay in the presence of neutrino mixing~\cite{Aluw,Blasone:2018czm,Cozzella:2018qew}. This subject was firstly addressed in Ref.~\cite{Aluw}, where the authors find a discrepancy between the two decay rates, concluding that the problem must be solved experimentally. However, in Ref.~\cite{Blasone:2018czm}, it was shown that this contradiction is connected with the incorrect choice of neutrino mass eigenstates as asymptotic states in the comoving frame\footnote{The authors of Ref.~\cite{Aluw} motivate this choice by the requirement of KMS thermal condition for the accelerated neutrino vacuum. However, in Ref.~\cite{Blasone:2017nbf}, it has been shown that the thermality of Unruh radiation for mixed fields is not violated, at least within the first order approximation we deal with in Ref.~\cite{Blasone:2018czm}. Furthermore, it has been recently pointed out that the KMS condition is not necessary at all for the Unruh effect to be present in QFT~\cite{Carballo-Rubio:2018zll}.}: if one instead adopts flavor eigenstates, the two decay rates perfectly agree~\cite{Blasone:2018czm}, at least within the approximation in which such states are described by the usual Pontecorvo ones~\cite{Bilenky:1978nj}.

Subsequently, another work dealing with this problem appeared~\cite{Cozzella:2018qew}, in which the authors conclude that no contradiction arises at all in connection with neutrino mixing. In their derivation, they claim that flavor states can only
be defined ``phenomenologically''~\cite{giunti}: in this way, their calculation basically reduces to the one of Ref.~\cite{Suzuki:2002xg}, with the difference that they use neutrinos with definite masses in the weak interaction vertices. It is clear, however, that a problem of exquisitely  theoretical nature, such as the possible violation of General Covariance and/or thermality of Unruh effect, cannot be correctly addressed without using a formalism fully consistent with the general theoretical framework for that phenomenon, namely the Standard Model.

Actually, the construction of flavor states for mixed particles has been carried out in a series of papers~\cite{Blasone:1995zc,Blasone:2001du,Blasone:2001qa}, where they have been rigorously defined as eigenstates of the flavor charge operators, obtained as usual via Noether's theorem:
\begin{eqnarray}
&&Q_\ell=\int d^3 x \,\bar{\nu}_\ell(x)\nu_\ell(x),\,\quad\ell=e,\mu,\\[2mm]
&&Q_\ell|0(\theta)\rangle_{e,\mu}=0, \quad Q_\ell|\nu_{k,\ell}\rangle=|\nu_{k,\ell}\rangle, \quad Q_\ell|\bar\nu_{k,\ell}\rangle=-|\bar\nu_{k,\ell}\rangle.
\label{nustates}
\end{eqnarray}
Here $\theta$ is the mixing angle and $|0(\theta)\rangle_{e,\mu}$ is the vacuum for definite flavor fields. The key point is that the transformation connecting neutrino annihilation operators with definite flavors to those with definite masses is not simply a rotation, but contains a Bogoliubov transformation~\cite{Blasone:2016isw}. A consequence of this fact is that flavor and mass representations are unitarily inequivalent in the infinite volume limit: 
\be
\qquad\lim_{V \rightarrow \infty}\; _{1,2}\langle0|0(\theta)\rangle_{e,\mu}=0, 
\ee  
where $|0\rangle_{1,2}$ is the vacuum for free fields $\nu_1$, $\nu_2$ with mass $m_1$, $m_2$.
 Oscillation formulas derived within this formalism~\cite{Blasone:2002wp} exhibit corrections with respect to the usual ones~\cite{Bilenky:1978nj} but, in the relativistic limit, Pontecorvo formulas and states are shown to be approximately recovered.

In this paper, we present the main points of the analysis of the inverse $\beta$ decay with neutrino mixing, as given in Ref.~\cite{Blasone:2018czm}. It is important to stress that the use made there of Pontecorvo states instead of the exact neutrino flavor states Eq.~(\ref{nustates}) is justified by the fact that calculations of Ref.~\cite{Blasone:2018czm} are performed in an approximation such that the result is insensitive to the choice between these two sets of states (see Appendix for more details).

Here, for simplicity we work in a two-dimensional spacetime and in natural units $\hbar=c=1$, with the metric signature $\eta_{\mu\nu}\, =\, {\rm diag}(+1,-1)$.

\section{Evaluation of the inverse $\beta$ decay rate} 
\label{sect1}
In this Section we investigate the decay of accelerated protons. Following Ref.~\cite{Matsas:1999jx},  calculations will be performed both in the inertial and comoving frame. 

In the approximation of small acceleration $a\ll M_{W^\pm}, M_{Z^0}$, by defining the Hermitian monopole $\hat{q}(\tau)$ as in Ref.~\cite{Birrell:1982ix}, the vector current describing the uniformly accelerated nucleons can be written as
\be
\hat{J}_{\mu}\,=\,\hat{q}(\tau)\hspace{0.2mm}u_{\mu}\delta \left(u-a^{-1}\right),
\ee 
where $u=a^{-1}=\mathrm{const}$ is the spatial Rindler coordinate representing the world line of nucleons with proper acceleration $a$, $\tau=v/a$ is the proper time and $v$ is the Rindler time coordinate. The nucleons' four velocity $u^{\mu}$ is defined by $u^{\mu}=(a, 0)$ and $u^{\mu}=(\sqrt{a^2t^2+1}, at)$ in Rindler and Minkowski coordinates, respectively\footnote{\,The Rindler coordinates $(v, u)$  are related with the Minkowski coordinates $(t, z)$ by: $t=u\sinh{v}$, $z=u\cosh{v}$.}.

Using Fermi theory, we can express the coupling of the electron $\hat{\Psi}_{e}$ and neutrino $\hat\Psi_{\nu}$ fields to the hadronic current $\hat{J}_{\mu}$ as follows
\be
\label{eqn:Fermiaction}
\hat{S}_{I}\ =\ \hspace{-0.5mm}\int d^{2}x\hspace{0.2mm}\sqrt{-g}\hspace{0.3mm}\hat{J}_{\mu}\hspace{-0,3mm}\left(\hat{\overline{\Psi}}_{\nu_e}\gamma^{\mu}\hat{\Psi}_{e}\ +\ \hat{\overline{\Psi}}_{e}\gamma^{\mu}\hat{\Psi}_{\nu_e}\right),
\ee
where $g$ is the determinant of the metric and $\gamma^\mu$ are the gamma matrices in the Dirac representation~\cite{Itzykson}.

\subsection{Inertial frame}
\label{subsect1}
\begin{figure}[t]
\centering\resizebox{8cm}{!}{\includegraphics{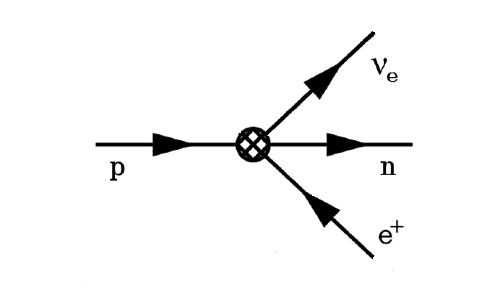}}
\caption{\small{Proton decay in the inertial frame.}
}
\label{figure:Inertial}
\end{figure}

In the inertial frame, the accelerated proton decays into a neutron by emitting a positron and a neutrino, as shown in Fig.~\ref{figure:Inertial}.
Denoting by $m_{e(\nu)}$ the mass of the electron (neutrino) field and by $k_{e(\nu)}$, $\omega_{e(\nu)}=\sqrt{m^2_{e(\nu)}+k^2_{e(\nu)}}$, $\sigma_{e(\nu)}$ the momentum, frequency and polarization of the modes in Minkowski space, respectively, the tree-level transition amplitude for the decay process takes the form
\begin{equation}
\mathcal{A}^{p\rightarrow n}_{in}\ \equiv\ \left\langle n\right|\otimes\left\langle e_{k_{e}\sigma_{e}}^{+},\nu_{k_{\nu}\sigma_{\nu}}\right|\hat{S}_{I}\left|0\right\rangle\otimes\left|p\right\rangle\ =\ \frac{G_F}{2\pi}\,\mathcal{I}_{\sigma_\nu\sigma_e}(\omega_\nu, \omega_e),
\label{tramp}
\end{equation}
where we used the standard expansion of Dirac fields in Minkowski space~\cite{Matsas:1999jx} and we introduced the shorthand notation
\begin{eqnarray}
\label{I}
\nonumber
\mathcal{I}_{\sigma_\nu\sigma_e}(\omega_\nu, \omega_e)&\,\equiv\,&\hspace{-1.1mm}\int_{-\infty}^{+\infty}\hspace{-2.3mm}d\tau\, e^{i\left[\Delta m\tau+a^{-1}\left(\omega_e+\omega_{\nu}\right)\sinh a\tau-a^{-1}\left(k_e+k_{\nu}\right)\cosh a\tau\right]}\\[2mm]
&&\times\left[\cosh a\tau\, {g}^{(+\omega_{\nu})\hspace{0.1mm}\dagger}_{k_{\nu}\sigma_{\nu}}\hspace{0.mm}g^{(-\omega_e)}_{-k_e-\sigma_e}\hspace{-1mm}\ +\ \sinh{a\tau}\,\bar{g}^{(+\omega_{\nu})}_{k_{\nu}\sigma_{\nu}}\,\gamma^3\,g^{(-\omega_e)}_{-k_e-\sigma_e}\right]\hspace{-0.7mm}.
\end{eqnarray}
Here, $g_{k\sigma}^{(\pm\omega)}$ are the Dirac modes in Minkowski space up to the exponential factor $e^{(\mp\omega t+kz)}/\sqrt{2\pi}$~\cite{Matsas:1999jx} and  $\Delta m$ is the difference of the nucleon masses.

The differential and total transition rates are defined as 
\be
\label{ttr}
\frac{d^{2}\mathcal{P}_{in}^{p\rightarrow n}}{dk_{\nu}dk_{e}}\ =\ \sum_{\sigma_{\nu}=\pm}\sum_{\sigma_{e}=\pm}\left|\mathcal{A}_{in}^{p\rightarrow n}\right|^{2},\qquad\quad\Gamma_{in}^{p\rightarrow n}\ =\ \mathcal{P}_{in}^{p\rightarrow n}/T,
\ee
where $T=\int_{-\infty}^{+\infty}ds$ is the nucleon proper time. The calculation of $\Gamma_{in}^{p\rightarrow n}$ gives
\begin{equation}
\Gamma_{in}^{p\rightarrow n}=\frac{4\,G_{F}^{2}}{a\,\pi^{2}\,e^{\pi\Delta m/a}}\int_{0}^{\infty}d{k}_{e}\int_{0}^{\infty}d{k}_{\nu}\left\{K_{2i\Delta m/a}\left[2\left(\frac{\Omega}{a}\right)\right]\,+\,\frac{m_{e}m_{\nu}}{\omega_{e}\omega_{\nu}}\,\mathrm{Re}\left\{K_{2i\Delta m/a+2}\left[2\left(\frac{\Omega}{a}\right)\right]\right\}\right\},
\label{finalinertial}
\end{equation}
where $\Omega\equiv{\omega}_{e}+{\omega}_{\nu}$. Details can be found in Refs.~\cite{Blasone:2018czm, Suzuki:2002xg} for a more general four-dimensional treatment.

\subsection{Comoving frame }
\begin{figure}[h]
\centering\resizebox{15cm}{!}{\includegraphics{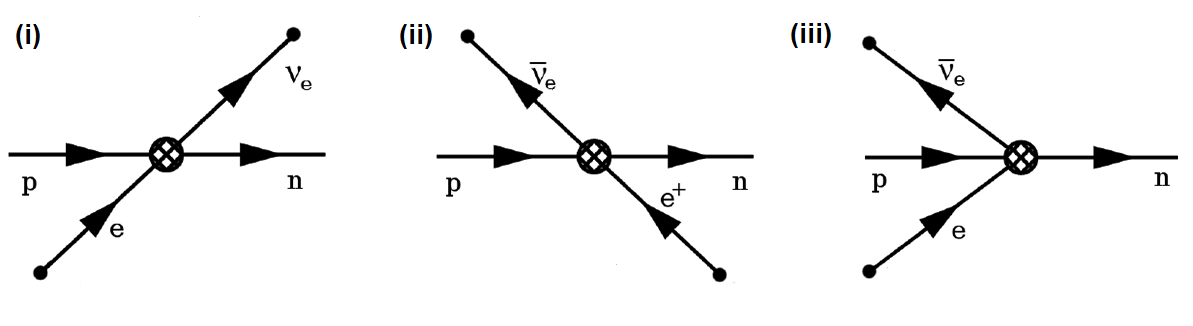}}
\caption{\small{Proton decay in the comoving frame.}
}
\label{threeprocesses}
\end{figure}

Let us now analyze the proton decay in the comoving frame. In this case, the process is allowed by the absorption (emission) of $e^-$ and $\bar{\nu}_e$ ($e^+$ and ${\nu}_e$) from (to) the Unruh thermal bath~\cite{Unruh:1976db}. 
Formally, the evaluation of the transition rate can be performed in the same way as the inertial frame. Here, however,  the relevant processes are those shown in Fig.~\ref{threeprocesses}.

We begin by considering the process $(i)$. Using the standard Rindler expansion for  fields in a uniformly accelerated frame~\cite{Matsas:1999jx}, a straightforward calculation gives
\begin{equation}
\label{first}
\mathcal{A}^{p\rightarrow n}_{(i)}\ \equiv\ \left\langle n\right|\otimes\langle\nu_{\omega_\nu\hspace{0.2mm}\sigma_{\nu}}|\hspace{0.2mm}\hat{S}_{I}\hspace{0.2mm}|e^{-}_{\omega_{e^-}\hspace{0.2mm}\sigma_{e^-}}\rangle\otimes\left|p\right\rangle\ =\ \frac{G_F}{a}\mathcal{J}_{\sigma_\nu\sigma_e}(\omega_\nu, \omega_e),
\end{equation}
where $\hat{S}_I$ is the Fermi action Eq.~(\ref{eqn:Fermiaction}) with $\gamma^\mu$ replaced by the Rindler gamma matrices $\gamma^{\mu}_R={(e_a)}^{\mu}\hspace{0.2mm}\gamma^a$,  ${(e_a)}^{\mu}$ being the tetrads, and 
\be
\label{eqn:j}
\mathcal{J}_{\sigma_\nu\sigma_e}(\omega_\nu, \omega_e)=\int_{-\infty}^{+\infty}\hspace{-1mm}dv\, e^{i\left[\Delta m+\omega_\nu-\omega_e\right]v/a}\,{h}_{\omega_\nu\sigma_\nu}^{(m_\nu)\dagger}\, h_{\omega_e\sigma_e}^{(m_e)}.
\ee
Here, we denoted by $h_{\omega\sigma}^{(m)}$ the Rindler modes up to the exponential factor $e^{-i\omega v/a}$~\cite{Matsas:1999jx}.

Since the probability that the proton absorbs (emits) a particle from (to) the Unruh thermal bath is $n_{F}(\omega)\,=\,\frac{1}{1+e^{2\pi\omega/a}}$ $\left(1-n_{F}(\omega)\right)$~\cite{Unruh:1976db}, the differential transition rate for the process $(i)$ becomes
\begin{eqnarray}
\label{eqn:diftransraterind}
\frac{1}{T}\frac{d^{2}\mathcal{P}^{p\rightarrow n}_{(i)}}{d\omega_{\nu}d\omega_e}=\frac{1}{T}\sum_{\sigma_{\nu},\sigma_e}\left|\mathcal{A}^{p\rightarrow n}_{(i)}\right|^{2}n_{F}(\omega_{e})[1-n_{F}(\omega_\nu)].
\end{eqnarray}
Analogous calculations for the processes $(ii)$ and $(iii)$ lead to the following total decay rate:
\be\label{finalacc}
\Gamma_{acc}^{p\rightarrow n}\ \equiv\ \Gamma_{(i)}^{p\rightarrow n}+\Gamma_{(ii)}^{p\rightarrow n}+\Gamma_{(iii)}^{p\rightarrow n}\ =\ \frac{4\hspace{0.2mm}G_{F}^{2}\hspace{0.2mm}m_{\nu}\hspace{0.2mm}m_{e}}{\pi^3\hspace{0.2mm}a^{2}\hspace{0.2mm}e^{\pi\Delta m/a}}\int_{-\infty}^{+\infty}\hspace{-2mm}d\omega_e\,\mathcal{R}^{2}(\omega_e),
\ee 
where
\be
\label{R}
\mathcal{R}(\omega_e)\ \equiv\ \mathrm{Re}\left[K_{i(\omega_{e}-\Delta m)/a-1/2}(m_{\nu}/a)K_{i\omega_{e}/a+1/2}(m_{e}/a)\right].
\ee
As in the inertial case, the explicit derivation of $\Gamma^{p\rightarrow n}_{acc}$ is rather awkward (see Refs.~\cite{Blasone:2018czm,Suzuki:2002xg} for details). The crucial point, however, is that the decay rates in the two frames perfectly agree with each other.

\section{Inverse $\beta$ decay with mixed neutrinos}
In the previous Section, we have considered  electron neutrino as a fundamental field, acting on $|\nu_e\rangle$ as a free-field like operator. In the Standard Model (SM), however, it is well-known that neutrinos interact weakly with other particles in flavor eigenstates $|\nu_{\ell}\rangle$ ($\ell\,=\,e, \mu$) that are coherent superpositions of mass eigenstates\footnote{For the sake of simplicity, we consider a two flavor model.} $|\nu_i\rangle$ ($i\ =\ 1, 2$).  The relation between these two sets of states is given by
\be
\begin{pmatrix}
\label{Pontec}
|\nu_e\rangle_P\\
|\nu_\mu\rangle_P
\end{pmatrix}\ =\ U(\theta)\begin{pmatrix}
|\nu_1\rangle\\
|\nu_2\rangle
\end{pmatrix}
\equiv
 \begin{pmatrix}
\cos\theta&\sin\theta\\
-\sin\theta&\cos\theta
\end{pmatrix}
\begin{pmatrix}
|\nu_1\rangle\\
|\nu_2\rangle
\end{pmatrix},
\ee
where $U(\theta)$ is the Pontecorvo unitary mixing matrix~\cite{Bilenky:1978nj}. We have introduced a subscript $P$ to distinguish the Pontecorvo flavor states above defined from the ones mentioned in the Introduction, which are the eigenstates of flavor charge (see also Appendix).

Let us then analyze how calculations of previous Section merge with Pontecorvo transformation Eq.~(\ref{Pontec}).

\subsection{Inertial frame}

We start by rotating both neutrino fields and states in Eq.~(\ref{tramp}) according to Eq.~(\ref{Pontec}), obtaining
\begin{equation}\label{ntc}
\mathcal{A}_{in}^{p\rightarrow n}\ =\ \frac{G_F}{2\pi}\left[\cos^2\theta\, \mathcal{I}_{\sigma\sigma_e}(\omega_{\nu_1},\omega_e)+ \sin^2\theta\, \mathcal{I}_{\sigma\sigma_e}(\omega_{\nu_2},\omega_e)\right],
\end{equation}
where $\mathcal{I}_{\sigma\sigma_e}(\omega_{\nu_j},\omega_e)$, $j=1,2$, is defined as in Eq.~(\ref{I}) for each of the two mass eigenstates. Using Eq.~(\ref{ttr}), we then obtain the following expression for the total decay rate $\Gamma_{in}^{p\rightarrow n}$:
\be
\label{eqn:inertresultat}
\Gamma^{p\rightarrow n}_{in}\ =\ \cos^4\theta\, \Gamma^{p\rightarrow n}_{1}\ +\ \sin^4\theta\,\Gamma^{p\rightarrow n}_{2}\ +\ \cos^2\theta\sin^2\theta\,\Gamma^{p\rightarrow n}_{12},
\ee
with the simplified notation
\begin{eqnarray}
\label{integral}
\Gamma^{p\rightarrow n}_{j}\,& \equiv\,& \frac{1}{T}\sum_{\sigma,\sigma_e}\frac{{G_F}^2}{4\pi^2}\int^{+\infty}_{-\infty}\hspace{-1.5mm} dk\int^{+\infty}_{-\infty}\hspace{-1.5mm} dk_e\,{|\mathcal{I}_{\sigma\sigma_e}(\omega_{\nu_j},\omega_e)|}^2,\qquad j\,=\,1,2,\\[2mm]
\label{integral12}
\Gamma^{p\rightarrow n}_{12}\,& \equiv\,& \frac{1}{T}\sum_{\sigma,\sigma_e}\frac{{G_F}^2}{4\pi^2}\int^{+\infty}_{-\infty}\hspace{-1.5mm} dk\int^{+\infty}_{-\infty}\hspace{-1.5mm} dk_e\left[\mathcal{I}_{\sigma\sigma_e}(\omega_{\nu_1},\omega_e)\,\mathcal{I}_{\sigma\sigma_e}(\omega_{\nu_2},\omega_e)^*\ +\ \mathrm{c.c.}\right].
\end{eqnarray}

It is worth to note that, whilst the integrals $\Gamma^{p\rightarrow n}_{j}$ ($j=1,2$) can be solved analytically~\cite{Matsas:1999jx}, the treatment of the off-diagonal term $\Gamma^{p\rightarrow n}_{12}$ is absolutely non-trivial. A thorough analysis is discussed in Ref.~\cite{Blasone:2018czm}. 

We remark that the above off-diagonal term is completely absent in the analysis of Ref.~\cite{Cozzella:2018qew}, which is based on a phenomenological definition of flavor states~\cite{giunti}. Neverthless, in Refs.~\cite{giunti}, it is also argued that flavor states can be defined in the relativistic limit. Thus, one expects that the outcome of Ref.~\cite{Cozzella:2018qew} -- which is claimed to be exact -- would reproduce our approximate result in the relativistic limit (where Pontecorvo states are known to be well-defined and indeed describe the phenomenology of observed neutrino oscillations) and for small neutrino mass difference. However, as remarked, this does not happen.

\subsection{Comoving frame}
\label{sect2}
Let us now turn the attention to the comoving frame. As it has been shown in Ref.~\cite{Blasone:2018czm}, assuming asymptotic neutrinos to be mass eigenstates would inevitably lead to a disagreement between the decay rates in two frames -- a result that is incompatible with the General Covariance of the underlying formalism.

Letting ourselves be guided by the lighthouse of  the General Covariance, we require the asymptotic neutrino states in the comoving frame to be flavor eigenstates (see footnote 1 for a discussion about the approach of Ref.~\cite{Aluw} to this point). 
By referring to the process $(i)$ in Fig.~(\ref{threeprocesses}), the transition amplitude Eq.~(\ref{first}) becomes
\be
\mathcal{A}_{(i)}^{p\rightarrow n}\ =\ \frac{G_{F}}{ a}\left[\cos^2\theta\mathcal{J}^{(1)}_{\sigma\sigma_e}(\omega, \omega_e)\ +\ \sin^2\theta\mathcal{J}^{(2)}_{\sigma\sigma_e}(\omega, \omega_e)\right],
\ee
where $\mathcal{J}^{(j)}_{\sigma\sigma_e}(\omega, \omega_e)$ ($j=1,2$) is defined as in Eq.~(\ref{eqn:j}) for each of the two neutrino mass eigenstates. Analogous considerations for the processes $(ii)$ and $(iii)$ finally lead to the following expression for  the total transition rate:
\begin{eqnarray}
\label{gammacc}
\Gamma_{acc}^{p\rightarrow n}\, \equiv\, \Gamma_{(i)}^{p\rightarrow n}\, +\, \Gamma_{(ii)}^{p\rightarrow n}\,+\,\Gamma_{(iii)}^{p\rightarrow n}\, =\,\cos^4\theta\, \widetilde\Gamma^{p\rightarrow n}_{1}\ +\ \sin^4\theta\,\widetilde\Gamma^{p\rightarrow n}_{2}\ +\ \cos^2\theta\sin^2\theta\,\widetilde\Gamma^{p\rightarrow n}_{12},
\end{eqnarray}
where $\widetilde\Gamma^{p\rightarrow n}_{j}$ ($j=1,2$) is defined as
\begin{equation}
\label{integralbis}
\widetilde\Gamma^{p\rightarrow n}_{j}\ \equiv\ \frac{4G_{F}^{2}m_{e}\hspace{0.2mm}m_{\nu_j}}{\pi^3 a^{2}e^{\pi\Delta m/a}}\int_{-\infty}^{+\infty}\hspace{-1.5mm}d\omega_e\hspace{0.4mm}R_{j}^{2}(\omega_e),\qquad j=1,2,
\end{equation}
and
\begin{equation}
\label{int1212}
\widetilde\Gamma^{p\rightarrow n}_{12}\ =\ \frac{8\hspace{0.2mm}G_{F}^{2}m_{e}\sqrt{m_{\nu_1}m_{\nu_2}}}{\pi^3 a^{2}e^{\pi\Delta m/a}}\int_{-\infty}^{+\infty}\hspace{-1.5mm}d\omega_e\hspace{0.4mm}R_{1}(\omega_e)\hspace{0.2mm}R_{2}(\omega_e).
\end{equation}
By comparing Eqs.~(\ref{eqn:inertresultat}),~(\ref{gammacc}) and exploiting the following equality~\cite{Matsas:1999jx}
\be
\label{equality}
\Gamma^{p\rightarrow n}_{j}\ =\widetilde\Gamma^{p\rightarrow n}_{j}\qquad j\,=\,1,2,
\ee
we thus realize that the inertial and comoving results would match, provided that the off-diagonal terms $\Gamma^{p\rightarrow n}_{12}$ (Eq.~(\ref{integral12}))  and $\widetilde\Gamma^{p\rightarrow n}_{12}$ (Eq.~(\ref{int1212})) coincide.  It is quite difficult to 
draw a final conclusion at this stage. From a preliminary analysis carried out in the limit of small neutrino mass difference $\frac{\delta m}{m_{\nu_1}}\,\ll\, 1$~\cite{Blasone:2018czm}, however, it has been shown that $\Gamma^{p\rightarrow n}_{12}=\widetilde\Gamma^{p\rightarrow n}_{12}$  to the leading order in $\frac{\delta m}{m_{\nu_1}}$. Asserting whether this equality holds exactly is not a foregone conclusion, thus leaving the problem of the inverse $\beta$ decay in the context of mixing open even when taking neutrino asymptotic states to be flavor eigenstates. A detailed discussion of this issue is addressed in the last Section.

\section{Comments and Conclusions}
In this paper we have analyzed the r\^ole of neutrino mixing in the context of the inverse $\beta$ decay of accelerated protons. Working in the approximation of small neutrino mass difference, we have shown that, in order for the two decay rates to coincide (and thus General Covariance to hold), asymptotic neutrino states must be taken as flavor eigenstates rather than mass eigenstates. In our calculations, we have employed the usual Pontecorvo states, since, in the considered limit, they well approximate the exact expression of flavor eigenstates Eq.~(\ref{nustates}) (see also Appendix).

Further investigation is inevitably required for the understanding of what happens beyond such an approximation. In view of this, two paths need to be considered. On the one hand, one may attempt to  keep on using Pontecorvo states. In this case, however, we suspect that the contradiction would not be solved, since Pontecorvo transformations Eq.~(\ref{Pontec}) are not consistent with the Quantum Field Theory formalism~\cite{{Blasone:1995zc}}. On the other hand, the adoption of exact neutrino flavor states Eq.~(\ref{nustates}) shall be pursued. Nevertheless, also in this scenario two possibilities should be contemplated.  If the two decay rates  coincide, then the discrepancy arising in Ref.~\cite{Aluw} would  be solved at a purely theoretical level. Instead, if the contradiction remains, one could envisage two potential sources of the problem (or combination thereof):
\begin{itemize}
\item[1.]Neutrino mixing in the context of the Standard Model is at odds with general covariance; 
\item[2.]Unruh effect should be somehow modified when neutrino mixing is taken into account (for example, violating the thermality of vacuum state~\cite{Blasone:2017nbf}).
\end{itemize}

It is an interesting question, and object of future work, to investigate in detail where the source of the inconsistency resides. In line with Refs.~\cite{Blasone:2017nbf}, however, we tend to regard the second option as the correct one.

\appendix
\section*{Appendix}
In this Appendix, we first review some elements of the quantization of mixed neutrino fields
and then we show that the use of Pontecorvo states in Ref.~\cite{Blasone:2018czm} is justified within the approximations there used, i.e. small neutrino mass difference and vanishing neutrino mass. 

The vacuum for definite flavor neutrinos $|0\rangle_{e,\mu}$ is expressed in terms of the vacuum for definite mass neutrinos $|0\rangle_{1,2}$ by~\cite{Blasone:1995zc}
\begin{eqnarray}\nonumber
|0\rangle_{e,\mu}&=&\prod_{k,\sigma}\Bigl[\left(1-\sin^2\theta|V_k|^2\right)-\varepsilon^\sigma\sin\theta\cos\theta V_k\left(A_k^\sigma+B_k^\sigma\right)\\[2mm]
&&+\varepsilon^\sigma\sin^2\theta\left(U_k^*\,C_k^\sigma-U_k\,D_k^\sigma\right)+\sin^2\theta|V_k|^2\,A_k^\sigma\,B_k^\sigma\Bigr]\,|0\rangle_{1,2}\,,
\label{vacua}
\end{eqnarray}
where $\varepsilon^\sigma=(-1)^\sigma$ and
\be
A_k^\sigma\equiv b_{k,1}^{\sigma\,\dagger}\,d_{-k,2}^{\sigma\,\dagger},\quad B_k^\sigma\equiv b_{k,2}^{\sigma\,\dagger}\,d_{-k,1}^{\sigma\,\dagger},\quad C_k^\sigma\equiv b_{k,1}^{\sigma\,\dagger}\,d_{-k,1}^{\sigma\,\dagger},\quad D_k^\sigma\equiv b_{k,2}^{\sigma\,\dagger}\,d_{-k,2}^{\sigma\,\dagger},
\ee
with $b_{k,j}^\sigma$ $(d_{k,j}^\sigma)$, $j=1,2$ being the annihilators for neutrinos (antineutrinos) of mass $m_j$, momentum $k$ and polarization $\sigma$. These operators are related to the corresponding annihilators and creators for neutrinos with definite flavor according to 
\be\label{bogol}
b_{k,e}^\sigma=\cos\theta\,b_{k,1}^\sigma+\sin\theta\left(U_k^*\,b_{k,2}^\sigma+\varepsilon^\sigma\,V_k\,d_{-k,2}^{\sigma\,\dagger}\right),\qquad b_{k,e}^\sigma|0\rangle_{e,\mu}=0,
\ee
and similar for the other operators. The above relation is the combination of a rotation and a Bogoliubov transformation. The Bogoliubov coefficients are defined as
\begin{eqnarray}
\label{Vk1}  \bogu=u_{k,2}^{\sigma\,\dagger}\,u_{k,1}^{\sigma}=v_{-k,1}^{\sigma\,\dagger} v_{-k,2}^{\sigma},\;\;\;\;\;\;\;\; \bogv=\varepsilon^\sigma u_{k,1}^{\sigma\,\dagger}\,v_{-k,2}^{\sigma}=-\varepsilon^\sigma u_{k,2}^{\sigma\,\dagger}\,v_{-k,1}^{\sigma},
\end{eqnarray}
where $u_{k,i}^\sigma$ ($v_{-k,i}^\sigma$) are the field modes for fermions (antifermions). By explicit calculation, it is possible to show that\footnote{Note that our results are consistent with the ones of Ref.~\cite{Blasone:1995zc}, although in that case calculations are performed in four-dimensions.  }~\cite{Blasone:1995zc}
\be
\label{Vk1}  \bogu=|\bogu|\;e^{i(\omega_{\nu_2}-\omega_{\nu_1})t},\;\;\;\;\;\;\;\; \bogv=|\bogv|\;e^{i(\omega_{\nu_2}+\omega_{\nu_1})t},
\ee
\begin{eqnarray}
\label{eqn:Uk} |\bogu|&=&\left(\frac{\omega_{\nu_1}+m_{1}}{2\omega_{\nu_1}}\right)^{\frac{1}{2}}
\left(\frac{\omega_{\nu_2}+m_{2}}{2\omega_{\nu_2}}\right)^{\frac{1}{2}}
\left(1+\frac{{
k}^{2}}{(\omega_{\nu_1}+m_{1})(\omega_{\nu_2}+m_{2})}\right),
\\[2mm]
\label{eqn:Vk}|\bogv|&=&\left(\frac{\omega_{\nu_1}+m_{1}}{2\omega_{\nu_1}}\right)^{\frac{1}{2}}
\left(\frac{\omega_{\nu_2}+m_{2}}{2\omega_{\nu_2}}\right)^{\frac{1}{2}}
\left(\frac{k}{(\omega_{\nu_2}+m_{2})}-\frac{k}{(\omega_{\nu_1}+m_{1})}\right),
\end{eqnarray}
with
\begin{eqnarray}
|\bogu|^{2}+|\bogv|^{2}=1.
\label{eqn:circo}
\end{eqnarray}

Now, the one electron neutrino state is given by
\be\label{oen}
|\nu_{k,e}^\sigma\rangle\,\equiv\, b_{k,e}^{\sigma\,\dagger}\,|0\rangle_{e,\mu}.
\ee
Note that, in the calculations of the decay rate in the inertial frame Eq.~(\ref{eqn:inertresultat}), the following contributions concerning the neutrino sector appear in the form (integration omitted)~\cite{Blasone:2018czm}
\be
|{}_P\langle \nu_e|\widehat{\overline{\Psi}}_{\nu_e}|0\rangle_{1,2}|^2=\cos^4\theta+\sin^4\theta+2\cos^2\theta\sin^2\theta|U_{k_{\nu}}|,
\ee
where $|\nu_e\rangle_P$ is the Pontecorvo state introduced in Eq.~(\ref{Pontec}). 

In the approximation of Ref.~\cite{Blasone:2018czm}, we have
\be
|U_{k_\nu}|\approx 1-\frac{\delta m^2}{8k_\nu^2}+O(\delta m^4).
\ee
To the leading order, we then obtain
\be
|{}_P\langle \nu_e|\hat{\overline{\Psi}}_{\nu_e}|0\rangle_{1,2}|^2\approx 1,
\ee
that is the same result we would obtain using the exact neutrino flavor state Eq.~(\ref{oen}) and the flavor vacuum Eq.~(\ref{vacua}) instead of Pontecorvo states and mass vacuum, thus justifying the employment of the Pontecorvo states in our calculations.

\end{document}